\newif\iffinal
\finaltrue
\newif\iffull
\fulltrue

\iffinal
\else
\fulltrue
\fi

\documentclass[submission,copyright,creativecommons]{eptcs}
\providecommand{\event}{SYNT 2014} 
\usepackage{breakurl}             

\usepackage{amsmath,amssymb,amsfonts,amsthm}
\usepackage{pstricks,pst-node,wrapfig}
\usepackage{temporal}
\usepackage{xspace}
\usepackage{booktabs}
\usepackage{paralist}
\usepackage{listings} 
\makeatletter    
\let\note\@undefined                      
\let\endnote\@undefined  
\makeatother     %
\iffinal
  \usepackage[finalold]{trackchanges} 
\else
  \usepackage[margins]{trackchanges}
\fi
\usepackage{soul}
\usepackage{enumitem}
\usepackage{color,xcolor}
\usepackage{graphicx}
\usepackage{caption}
\usepackage{hyperref} 

\newtheorem{corollary}{Corollary}

\newtheorem{observation}{Observation}
\newtheorem{theorem}{Theorem}

\iffinal
\newcommand{\comment}[1]{}
\else
\newcommand{\comment}[1]{{\footnotesize {\color{black!70} #1}}}
\fi

\newcommand{\cupdot}{\mathbin{\dot{\cup}}}

\newcommand{\ambasignal}[1]{\textsc{#1}}
\newcommand{\ambasignali}[2][i]{\textsc{#2}[\textrm{#1}]}

\newcommand{\hgrant}{\ambasignal{hgrant}}
\newcommand{\hgranti}[1][i]{\ambasignali[#1]{hgrant}}
\newcommand{\hbusreq}{\ambasignal{hbusreq}}
\newcommand{\hbusreqi}[1][i]{\ambasignali[#1]{hbusreq}}
\newcommand{\hready}{\ambasignal{hready}}

\newcommand{\hlock}{\ambasignal{hlock}}
\newcommand{\hlocki}{\ambasignali[i]{hlock}}
\newcommand{\hmastlock}{\ambasignal{hmastlock}}
\newcommand{\hmastlocki}[1][i]{\ambasignali[#1]{hmastlock}}
\newcommand{\hburst}{\ambasignal{hburst}}

\newcommand{\hmaster}{\ambasignal{hmaster}}
\newcommand{\hmasteri}[1][i]{\ambasignali[#1]{hmaster}}
\newcommand{\hstart}{\ambasignal{start}}
\newcommand{\hstarti}[1][i]{\ambasignali[#1]{start}}
\newcommand{\hlocked}{\ambasignal{locked}}
\newcommand{\hlockedi}[1][i]{\ambasignali[#1]{locked}}
\newcommand{\hdecide}{\ambasignal{decide}}
\newcommand{\hdecidei}{\ambasignali[i]{decide}}
\newcommand{\tok}{\ambasignal{tok}}
\newcommand{\toki}{\ambasignali[i]{tok}}
\newcommand{\hincr}{\ambasignal{incr}}
\newcommand{\hburstfour}{\ambasignal{burst4}}
\newcommand{\hburstthree}{\ambasignal{burst3}}

\newcommand{\norequests}{\ambasignal{no\_req}}

\renewcommand{\|}{\mid}
\newcommand{\powerset}[1]{2^{#1}}

\newcommand{\specialcell}[2][c]{%
  \begin{tabular}[#1]{@{}c@{}}#2\end{tabular}}

\newcommand{\APproc}{{\mathrm{O}_{\mathrm{pr}}}}
\newcommand{\APsys}{{\mathrm{O}_{\mathrm{sys}}}}

\newcommand{\InputsProcAll}{{\mathrm{I}_{\mathrm{pr}}}}
\newcommand{\InputsProcLocal}{\mathrm{I}_{loc}}
\newcommand{\InputsProcGlobal}{\mathrm{I}_{glob}}
\newcommand{\InputsProc}{\InputsProcAll}
\newcommand{\InputsSys}{{\mathrm{I}_{\mathrm{sys}}}}

\newcommand{\Buchi}{B\"uchi\xspace}

\newcommand{\IndSet}{\bbN}

\newcommand{\token}{\mathsf{tok}}
\newcommand{\sendi}{\ambasignali[i]{send}}
\newcommand{\send}{\ambasignal{send}}

\newcommand{\FairSched}{FairSched}

\newcommand{\pring}{\ensuremath{\mathcal {R}}}

\newcommand{\impl}{\, \rightarrow \, }

\newcommand{\bbN}{\mathbb{N}}

\newcommand{\trcv}{\mathsf{rcv}}
\newcommand{\tsnd}{\mathsf{snd}}
\newcommand{\Locals}{Q}
\newcommand{\LocalsI}{\Locals_0}

\newcommand{\ActionsProc}{\Sigma_{\mathsf{pr}}}

\newcommand{\ActionsSys}{\Sigma_{\mathsf{sys}}}
\newcommand{\Trans}{\delta}
\newcommand{\LabelFun}{\lambda}

\newcommand{\trans}[3]{#1 \stackrel{\mathsf{#3}}{\rightarrow} #2}

\addeditor{RB}
\addeditor{SJ}
\addeditor{AK}

\newcommand{\ak}[1]{\note[AK]{#1}}
\newcommand{\sj}[1]{\note[SJ]{#1}}

\iffinal
\newcommand{\todo}[1]{}
\else
\newcommand{\todo}[1]{\hl{#1}}
\fi

\newcommand\LTL{\ensuremath{\mbox{\textsf{LTL}}}}
\newcommand\LTLmX{\ensuremath{\mbox{\textsf{LTL}}\backslash\textsf{X}}}

\gdef\dash---{\thinspace---\hskip.16667em\relax}
\gdef\endash--{\thinspace--\hskip.16667em\relax}

\newcommand\tidx\ell

\newcommand{\rank}{\rho}

\renewcommand{\paragraph}[1]{\smallskip \noindent \textbf{#1}}
\theoremstyle{definition}
\newtheorem*{example}{Example}

\title{Parameterized Synthesis Case Study: AMBA AHB\\ (extended version\footnote{The original version was submitted to SYNT 2014})}
\author{Roderick Bloem \qquad\qquad Swen Jacobs \qquad\qquad Ayrat Khalimov
\institute{Graz University of Technology, Austria}
\thanks{This work was supported by the Austrian Science Fund (FWF) under the RiSE National Research Network (S11406).}
\email{firstname.lastname@iaik.tugraz.at}
}

\def\titlerunning{Parameterized Synthesis Case Study: AMBA AHB}
\def\authorrunning{R. Bloem, S. Jacobs \& A. Khalimov}

\begin{document}

\maketitle

\begin{abstract}
We revisit the AMBA AHB case study that has been used as a benchmark for several reactive synthesis tools. Synthesizing AMBA AHB implementations that can serve a large number of masters is still a difficult problem. We demonstrate how to use parameterized synthesis in token rings to obtain an implementation for a component that serves a single master, and can be arranged in a ring of arbitrarily many components. We describe new tricks -- property decompositional synthesis, and direct encoding of simple GR(1) -- that together with previously described optimizations allowed us to synthesize a component model with 14 states in about 1 hour.
\end{abstract}


\section{Introduction}
\label{sec:intro}

By automatically generating correct implementations from a temporal logic 
specification, reactive synthesis tools can relieve system designers from 
tedious and error-prone tasks like low-level manual implementation and 
debugging. This great benefit comes at the cost of high computational complexity 
of synthesis, which makes synthesis of large systems an ambitious goal. 
For instance, Bloem et al.~\cite{Bloem12}
synthesize an arbiter for the ARM AMBA Advanced High
Performance Bus (AHB)~\cite{AMBASpec}. The results, obtained using 
RATSY~\cite{Bloem10c},
show that both the size of the implementation and the time for synthesis
increase steeply with the number of masters that the arbiter can
handle. This is unexpected, since an arbiter for $n+1$ masters is very 
similar to an arbiter for $n$ masters, and manual implementations grow only 
slightly with the number of masters. While recent results show that 
synthesis time and implementation size can be improved in standard LTL 
synthesis tools~\cite{Finkbeiner12,GodhalCH13}, the fundamental problem of increasing 
complexity with the number of masters can only be solved by adapting the 
synthesis approach itself. 

To this end, Jacobs and Bloem~\cite{Jacobs14} introduced the 
\emph{parameterized synthesis} approach. In parameterized synthesis, we 
synthesize a component 
implementation that can be used as a building block, replicating components 
to form a system that satisfies a specification for 
\emph{any} number of components. The approach is based on \emph{cutoff} 
results that 
have previously only been used to reduce the \emph{verification} of 
parameterized systems to systems of a small, fixed size. In particular, small 
cutoffs exist for token-ring networks, as shown by Emerson and 
Namjoshi~\cite{Emerso03}. These results can be extended to allow the 
reduction of the parameterized synthesis problem to a distributed 
synthesis problem with a fixed number of components, which can in turn be 
solved by a modification of the \emph{bounded synthesis} procedure of 
Finkbeiner and 
Schewe~\cite{Finkbeiner13}.
As experiments with the original, na\"ive implementation of parameterized 
synthesis revealed that only very small specifications could be handled, 
Khalimov et al.~\cite{TEPS} introduced a number of optimizations that 
improved runtimes for synthesis of token-ring systems by several orders of 
magnitude. In this paper, we will show how the resulting synthesis method 
can serve as the basis for synthesizing an implementation for the  
parameterized AMBA AHB specification.
\todo{must cite: Effective Synthesis of Asynchronous Systems from GR(1) Specifications.}

\smallskip \noindent {\bf Contributions.}
 We demonstrate how to synthesize a 
parameterized implementation of the AMBA AHB, with guaranteed correctness for 
any number of masters. To this end, we translate the LTL specification of the 
AMBA AHB (as found in~\cite{BarbaraThesis}) into a version that is suitable 
for parameterized synthesis in token rings, and address several challenges 
with respect to theoretical applicability and practical feasibility:
\begin{enumerate}
\item 
We show how to \emph{localize} global input and output signals, i.e., 
those that cannot be assigned to one particular master. This is necessary 
since our approach is based on the replication of components that act only on 
local information.
\item We introduce theoretical extensions of the cutoff results for 
token rings that support some of the features of the AMBA specification. 
This includes the handling (some but not all) of assumptions on local and global inputs, and of the fully asynchronous timing model (which includes the synchronous behavior intended in AMBA).
\item We show how to handle multiple process templates to support a \emph{special process} with properties different from other processes.
\item We describe further \emph{optimizations} that make synthesis feasible, in 
particular based on the insight that the AMBA protocol features three 
different types of accesses, and the control structures for these accesses 
can be synthesized (to some degree) independently.
\end{enumerate}
Finally, we report on our practical experience with parameterized synthesis of the AMBA protocol, pointing out weaknesses and open problems in current synthesis approaches.


\section{The AMBA Case Study}
\label{sec:amba}
ARM’s \emph{Advanced Microcontroller Bus Architecture} (AMBA)~\cite{AMBASpec} 
is a communication bus for a number of masters and
clients on a microchip. The most important part of AMBA is the \emph{Advanced 
High-performance Bus} (AHB), a system bus for the efficient connection of 
processors, memory, and devices. 

The bus arbiter is the critical part of the AHB, ensuring 
that only one master accesses the bus at any time. Masters send \hbusreq\ to 
the arbiter if they want access, and receive
\hgrant\ if they are allowed to access it.
Masters can also ask for different
kinds of \emph{locked transfers} that cannot be interrupted.

The exact arbitration 
protocol for AMBA is not specified. Our goal is to synthesize a protocol 
that guarantees safety and liveness properties. According to the 
specification, any device that is connected to the bus will react to an input 
with a delay of one time step. I.e., we are considering Moore machines. In 
the following, we introduce briefly which signals are used to
realize the arbiter of this bus for masters.

\paragraph{Requests and grants.} The identifier of the master which is 
currently active is stored in the $n+1$-bit signal \hmaster[$n$:0], 
with $n$ chosen such that the number of masters fit into $n+1$ bits.
To request the 
bus, master $i$ raises signal \hbusreqi. The arbiter decides who will be 
granted the bus next by raising signal \hgranti. When
the client raises \hready, the bus access starts at the next tick, and there 
is an update \hmaster[$n$:0] := i, where \hgranti\ is currently active.

\paragraph{Locks and bursts.} A master can
request a locked access by raising both \hbusreqi\ and \hlocki. In this case, 
the master additionally sets \hburst[1:0] to either \ambasignal{single} 
(single cycle access), \hburstfour\ (four cycle burst) or \hincr\ (unspecified 
length burst). For a \hburstfour\ access, the bus is
locked until the client has accepted 4 inputs from the master (each signaled by
raising \hready). In case of a \hincr\ access, the bus is
locked until \hbusreqi\ is lowered. The arbiter raises \hmastlock\ if the bus 
is currently locked.

\paragraph{\bf LTL specification.}
The original natural-language specification~\cite{AMBASpec} has been translated into a formal specification in the GR(1) fragment of LTL before~\cite{BarbaraThesis,Bloem12,GodhalCH13}. Figure~\ref{fig:AMBASpec} shows the environment assumptions and system guarantees from~\cite{BarbaraThesis} that will be the basis for our parameterized specification. The full specification is 
$ (A1 \land \ldots \land A4) \rightarrow (G1 \land \ldots \land G11)$.
\begin{figure}[t]
  \fbox{%
  \begin{minipage}{\textwidth}
  \input{amba-spec}
  \end{minipage}}
\caption{Formal specification of the AMBA AHB~\cite{BarbaraThesis}, in the GR(1) fragment of LTL.}
\label{fig:AMBASpec}
\end{figure}
%


\section{Definitions} 
\label{sec:preliminaries}

A \emph{labeled transition system (LTS) over sets $O$ of output variables and $I$ of input variables} is a tuple
$
(Q,Q_0,\Sigma,\delta,\lambda)
$
where 
 $Q$ is the set of {\em states}, 
 $Q_0 \subseteq Q$ is the set of {\em initial states}, 
 $\Sigma = \powerset{I}$ is the set of {\em inputs} (also called \emph{transition labels}), 
 $\delta \subseteq Q \times \Sigma \times Q$ is the {\em transition relation}, 
 and $\lambda:Q \to \powerset{O}$ is the \emph{output function} (also called {\em state-labeling} function). Variables from $O \cup I$ will be used as atomic propositions in our specifications.

\subsection{System Model}
\label{dfn:paramsys}\label{def:process_template}
In this section we define the token ring system -- the LTS that consists of replicated copies of a process connected in a uni-directional ring.
Transitions in a token ring system are either internal or synchronized (in which one process sends the token to the next process along the ring). 
The token starts in a non-deterministically chosen process.

Fix a set $\APproc$ of (local) \emph{output variables} that contain a distinguished output variable $\tsnd$, and a set $\InputsProc$ of (local) \emph{input variables} that contain a distinguished input variable $\trcv$.

\smallskip\noindent\textbf{Process Template $P$.} 
Let $\ActionsProc = \powerset{\InputsProc}$.
A {\em process template} $P$ is a LTS 
$(\Locals, \LocalsI, \ActionsProc, \Trans, \LabelFun)$ over $\APproc$:

\begin{enumerate}[label*=\roman*)]
\item 
The state set $\Locals$ is finite and can be partitioned into two non-empty disjoint sets: $\Locals = T \cupdot NT$. 
    States in $T$ are said to {\em have the token}.

\item
The initial state set is $Q_0  = \{\iota_t, \iota_n\}$ for some $\iota_t \in T, \iota_n \in NT$.

\item 
The output function $\LabelFun: \Locals \rightarrow \powerset{\APproc}$ satisfies that for every $q \in NT$, $\tsnd \not\in \LabelFun(q)$.

\item 
Let $\ActionsProc^\trcv = \{ i \in \ActionsProc \| \trcv \in i \}$, 
$\ActionsProc^{\neg \trcv} = \ActionsProc \setminus \Sigma^\trcv$,
$T^\tsnd = \{ q \in \Locals \| \tsnd \in \LabelFun(q) \}$,
$T^{\neg \tsnd} = T \setminus T^\tsnd$.
Then: 
$$\Trans \subseteq 
T^\tsnd \times \ActionsProc^{\neg \trcv}\times NT  ~~\cup~~
NT\times\ActionsProc^\trcv\times T   ~~\cup~~
NT\times\ActionsProc^{\neg \trcv}\times NT   ~~\cup~~
T^{\neg \tsnd}\times\ActionsProc^{\neg \trcv}\times T.$$

Also, $\Trans$ is non-terminating: 
for every $q \in NT$ and every ${\sf in} \in \ActionsProc$ there exists $\trans{q}{q'}{in} \in \Trans$; 
and for every $q \in T$ and every ${\sf in} \in \ActionsProc^{\neg \trcv}$ there exists $\trans{q}{q'}{in} \in \Trans$.

\item [$\dagger$)]
Given a fairness condition $A_{loc}$ over $\InputsProc \cup \APproc$,\footnote{Discussion of 
fairness conditions is deferred until Sect.~\ref{sec:semantics}.} 
we require that from any state $q$ with the token, under any input sequence satisfying $A_{loc}$, the 
process will reach a state $q'$ where it sends the token.
We call this requirement $\dagger$.

\end{enumerate}

\smallskip\noindent\textbf{Ring Topology $R$.}
A {\em ring} is a directed graph $R = (V,E)$, where the set of 
vertices is $V = \{ 1,\ldots,k\}$ for some $k \in \IndSet$, and the set of edges is 
$E = \{(i,i \oplus 1) \mid i\in V\}$. Vertices are called {\em process indices}.

\smallskip\noindent\textbf{Token-Ring System $P^R$.}
Fix a ring topology $R=(V,E)$.
Let $\InputsSys = (\InputsProcLocal \times V) \cupdot \InputsProcGlobal$ be the \emph{system input variables}, where local inputs $\InputsProcLocal$ and global inputs $\InputsProcGlobal$ are such that $\InputsProc = \InputsProcLocal \cupdot \InputsProcGlobal$.
Define $\ActionsSys = \powerset{\InputsSys}$.
For system input ${\sf in} \in \ActionsSys$, let ${\sf in}(v) = \{ i \in {\sf in} \| i\in \InputsProcLocal\times\{v\} \cup \InputsProcGlobal \}$ denote the input to process $v$ (including global inputs).
Let $\APsys = \APproc \times V$ be the \emph{system output variables}. 
For $(p,i)$ in $\APsys$ or in $\InputsSys\setminus \InputsProcGlobal$ we write $p_i$. 

Given a process template $P =  (\Locals, \LocalsI, \ActionsProc, \Trans, \LabelFun)$ over $\APproc$ and $\InputsProc$ and a token ring topology $R = (V,E)$, define the {\em token-ring system} $P^R$ as the finite LTS 
$(S,S_0,\ActionsSys,\Delta,\Lambda)$
over $\APsys$ and $\InputsSys$, where:
\begin{itemize}

\item 
The set $S$ of \emph{global states} is $Q^V$, i.e., all functions from $V$ to $Q$. If $s \in Q^V$ is a global state then $s(i)$ denotes the local state of the process with index $i$.

\item 
The set of \emph{global initial states} $S_0$ contains all $s_0 \in Q_0^V$ in which exactly one of the processes has the token.
    
\item 
The labeling $\Lambda(s) \subseteq \APsys$ for $s \in S$ is defined as follows: 
$p_i \in \Lambda(s)$ if and only if $p \in \LabelFun(s(i))$, for $p \in \APproc$ and $i \in V$.
  
\end{itemize}

Finally, define the {\em global transition relation} $\Delta$. 
In a {\em fully asynchronous token ring}, a subset of all processes can make a transition in each step of the system, i.e.,
$\Delta$ consists of the following set of transitions:

 \begin{itemize}
  \item 
  An {\em internal transition} is an element $(s,{\sf in},s')$ of $S \times \ActionsSys \times S$ for which there are process indices $M \subseteq V$ such that 
  \begin{enumerate}[label*=\roman*)]
    \item
    for all $v \in M$: $\tsnd \not\in \LabelFun(s(v))$ and $\trcv \not \in{\sf in}(v)$,

    \item 
    for all $v\in M$: $\trans{s(v)}{s'(v)}{in(v)}$ is a transition of $P$,

    \item 
    for all $u \in V \setminus M$: $s(u) = s'(u)$.
  \end{enumerate}

  \item
  A {\em token-passing transition}  is an element $(s,{\sf in},s')$ of $S \times \ActionsSys \times S$ for which there are process indices $M \subseteq V$ and two indices $v,w \in M$ such that $(v,w) \in E$ and

  \begin{enumerate}[label*=\roman*)]
    \item 
    $\tsnd \in \lambda(s(v))$, 
    and $\forall{u \in M\setminus \{v\}} : \tsnd \not\in \lambda(s(u))$ 
    --- i.e., only process $v$ sends the token,

    \item 
    $\trcv \in {\sf in}(w)$ and for all $u \in M \setminus \{w\}$: $\trcv \not \in{\sf in}(u)$ 
    --- i.e., only process $w$ receives the token,

    \item 
    for every $u\in M$: $\trans{s(u)}{s'(u)}{in(u)}$ is a transition of $P$,

    \item 
    for every $u \in V \setminus M$: $s'(u) = s(u)$. 
  \end{enumerate}
 
 \end{itemize}

Special cases of the fully asynchronous token ring are the \emph{synchronous token ring} and the \emph{interleaving token ring}. In a synchronous token ring, $M=V$ for internal and token-passing transitions. I.e., always all processes make a transition simultaneously. In an interleaving token ring, $M = \{v\}$ for some $v\in V$ for internal transitions, and $M=\{v,w\}$ for $(v,w)\in E$ for token-passing transitions. I.e., at each moment either \emph{exactly one} process makes an internal transition, or one process sends a token to the next process.

\smallskip\noindent\textbf{System Run.}
Fix a ring topology $R = (V,E)$.
A \emph{run of a token ring system} $P^{R}=(S,S_0,\ActionsSys,\Delta,\Lambda)$ over $\APsys$ and $\InputsSys$ is a finite or infinite sequence 
$x=(s_1,{\sf in}_1,M_1)(s_2,{\sf in}_2,M_2)\ldots$, where:
\begin{itemize}
  \item $s_1 \in S_0$, $s_k \in S$ and ${\sf in}_k \in \ActionsSys$ for any $k \le |x|$,
  \item for all $k < |x|: (s_k,{\sf in}_k,s_{k+1}) \in \Delta$,
  \item for all $k < |x|$: $M_k$ is the set of processes transiting in $(s_k,{\sf in}_k,s_{k+1})$ (see $M$ in the definition of $\Delta$).
\end{itemize}

\ak{what about maximality?}

\subsection{Parameterized Systems and Specifications}
\label{sec:semantics}
The \emph{parameterized ring} is the function $\pring: n \mapsto \pring(n)$, where $n \in \bbN$ and $\pring(n)$ is the ring with $n$ vertices. A \emph{parameterized token ring system} is a function $P^\pring: n \mapsto P^{\pring(n)}$, where $n\in\bbN$ and $P$ is a given process template.
When necessary to disambiguate we explicitly write `parameterized fully asynchronous token ring systems' or `parameterized interleaving token ring systems'.

A \emph{parameterized specification} is a sentence in indexed temporal logic, that is, a temporal logic formula with indexed variables and quantification over indices. Variables are from the set of output and input variables $\APproc \cup \InputsProc$, and indices refer to different copies of process templates. 
A parameterized token ring system $P^\pring$ \emph{satisfies} a parameterized specification $\phi$, written $P^\pring \models \phi$, iff $\forall n$: $P^{\pring(n)} \models \phi$. This definition assumes a fixed semantics for temporal logic formulas in labeled transition systems. Below, we introduce a slightly non-standard semantics as a modification of the (action-based) semantics in Emerson and Namjoshi~\cite{Emerso03} to the case of open systems.

\paragraph{Semantics for Open Systems.}
Emerson and Namjoshi~\cite{Emerso03} consider closed systems (i.e., without inputs), but with transitions labeled by \emph{actions} that can also be used in specifications. In the semantics defined below, we simulate an action {\sf a} by an input corresponding to {\sf a}. Furthermore, for defining when a given process satisfies a formula, we consider the projection of the run onto those points in time where the process actually makes a transition, just like actions are only considered when a transition fires.


In addition, we extend our semantics to the fully asynchronous timing model, which in particular includes the synchronous timing model that is needed for reasoning about the AMBA case study. \footnote{Note that in the synchronous timing model, our semantics is the same as the standard semantics (every process always makes a transition, so all inputs are considered), but we need the fully asynchronous case for the cutoff result in Thm.~\ref{thm:main-theorem}. Intuitively, in synchronous systems an implementation can count the number of global steps until it receives the token again, and therefore correctness of such implementations may depend on the size of the ring, making cutoff results impossible. Thus, systems that are correct in the synchronous but not in the fully asynchronous case are of limited interest to us.} This leads to additional problems: the natural ways to extend the semantics to properties of more than one process is to consider either the projection to those points in time where \emph{at least one} of the processes makes a step, or to those where \emph{all} processes make a step. Both cases are undesirable: in the first case, we also consider inputs that the processes cannot read, and in the second case the property will only be guaranteed at those points in time where all processes make a step together --- which is clearly undesirable, e.g., for a mutual exclusion property. Therefore, in properties that talk about more than one process, we do not allow input signals at all.

Fix a token ring system $P^{R}=(S,S_0,\ActionsSys,\Delta,\Lambda)$ over $\APsys$ and $\InputsSys$. 
We describe the semantics of parameterized properties 1-indexed properties over $\APsys$ and $\InputsSys$, and 2-indexed properties over $\APsys$.

\paragraph{Semantics of 1-indexed properties.}
\emph{1-indexed properties} are of the form $\forall{i}.\varphi(i)$, where $\varphi(i)$ is an \LTL\ formula over system variables that are indexed with $i$.

Fix a process index $j$. Given a run $(s_1,{\sf in}_1,M_1)(s_2,{\sf in}_2,M_2)\ldots$, consider the sub-sequence that contains exactly those $(s_k,{\sf in}_k,M_k)$ with $j \in M_k$. The \emph{local run of process $j$} is obtained by mapping each $(s_k,{\sf in}_k,M_k)$ in the sub-sequence to $(s_k(j),{\sf in}_k(j))$, where $s_k(j)$ is the local state of $j$ and ${\sf in}_k(j)$ are inputs to $j$.\sj{last part is repetition, may be removed} Then, \emph{a system run satisfies $\varphi(j)$} iff the local run of process $j$ satisfies $\varphi(j)$.
The latter satisfaction is defined in the usual way.
Note that since we consider only elements of the system run where process $j$ makes a step, we can use the next-time operator $\nextt$ (interpreted locally, cp.~\cite[Sect. 5.2]{TEPS}\cite[Sect. 2.5]{Emerso03}).

\begin{example}
Consider a typical 1-indexed property of an arbiter $\forall{i}.\always(r_i \impl \eventually g_i)$ (`for every process, every request should be finally granted').
In the semantics of 1-indexed properties described above, this property should be read as: `for every process, every request \emph{that has been seen by the process} should be finally granted'.
Another example: in the new semantics the property $\forall{i}.\always(r_i \impl \nextt g_i)$ should be read as `for every process, every request that has been seen by the process should be granted the next step'.
Notice that the environment cannot falsify the property by not scheduling the process.
\end{example}

\paragraph{Semantics of 2-indexed properties.} 
\emph{2-indexed properties} are of the form $\forall{i,j}.\varphi(i,j)$, where $\varphi(i,j)$ is an \LTLmX\ formula over output variables (and no input variables) that are indexed with $i$ or $j$.
Satisfaction for fixed process indices $i,j$ is defined in the standard way:
A system run $(s_1,{\sf in}_1,M_1)(s_2,{\sf in}_2,M_2)\ldots$ \emph{satisfies} $\varphi(i,j)$ iff the sequence $\Lambda(s_1)\Lambda(s_2)\ldots$ satisfies $\varphi(i,j)$.
I.e., we consider all elements of the system run, no matter if processes $i$ and $j$ make the transition or not.\sj{to save space, remove last sentence and example?}

\begin{example}
Consider a typical 2-indexed property of the mutual exclusion $\forall{i \neq j}.\always \neg (g_i \land g_j)$.
In the semantics of 2-indexed properties described above, the property should be read in a usual way: `it is never the case that two processes grant at the same time.
\end{example}

\paragraph{Semantics of ${\pforall_{\forall{i}.ass(i)} \varphi}$.}
Let $\varphi$ be a 1- or 2-indexed property as introduced above. 
\emph{A system satisfies $\pforall_{\forall{i}.ass(i)} \varphi$} iff any system run that satisfies $\forall{i}.ass(i)$ also satisfies $\varphi$, where satisfaction of 1-indexed $\forall{i}.ass(i)$ and of $\varphi$ as defined above.

\paragraph{Note on the Semantics of GR(1).}
The AMBA specification~\cite{BarbaraThesis} is defined in the GR(1) fragment of LTL, where the implication between assumptions and guarantees is usually interpreted with a special semantics~\cite{Klein10}. In this paper, we instead use the standard semantics for this implication.
\subsection{Parameterized Synthesis Problem}
\label{sec:synth-problem}

The \emph{parameterized synthesis problem} in token rings is: given a parameterized specification $\varphi$, find an implementation $P$ such that $P^\pring\models \varphi$. 
The problem is in general undecidable:

\begin{theorem}[\cite{Jacobs14}, Theorem 3.5]\label{thm:param-synth-is-undec}

  The parameterized synthesis problem of interleaving token rings with no global inputs is undecidable for specifications $\forall i,j. \pforall \varphi(i,j)$, where $\pforall \varphi(i,j)$ is an $\LTLmX$ formula over processes $i,j$.
\end{theorem}\ak{Over process $i,j$ outputs only, right?}
The proof reduces the undecidable problem of distributed synchronous synthesis \cite{Pnueli90} to distributed synthesis of an interleaving token ring of size 2, which implies undecidability of parameterized synthesis.

Note that Theorem \ref{thm:param-synth-is-undec} does not apply to specifications of the form $\pforall_{\forall{i}.ass(i)} \forall{j}.\varphi(j)$.
In fact, we can use the hub-abstraction technique (see \cite[Section 6]{TEPS}) to prove the following:

\begin{observation}
  The parameterized synthesis problem of token rings without global inputs is decidable for specifications $\forall{j}.\pforall_{\forall{i}.ass(i)} \varphi(j)$ where $ass(i)$ and $\varphi(i)$ are \LTL\ formulas over process $i$.\ak{what is the maximal bound on the size of a process?}
\end{observation}

\section{The Existing Parameterized Synthesis Approach}
\label{sec:approach}

The parameterized synthesis problem in token rings is in general undecidable, but Jacobs and Bloem~\cite{Jacobs14} have introduced a semi-decision procedure for the problem.
It is based on i) the cutoff results of \cite{Emerso03}, which state that model checking parameterized token rings is equivalent to model checking token rings of a cutoff size, and ii) the \emph{bounded synthesis} method~\cite{Finkbeiner13} that turns an undecidable synthesis problem (of synthesizing a token ring of fixed size) into a possibly infinite sequence of decidable synthesis problems (of synthesizing a token ring of fixed size in which a process implementation is not larger than the bound) by iterative bounding the size of process implementations.

\subsection{Cutoff Results in Token Rings}\label{sec:cutoffs}

A \emph{cutoff} for a parameterized specification $\varphi$ is a number $c \in \bbN$ s.t.
$P^{\pring(c)} \models \varphi 
~\iff~
\forall n \geq c:  P^{\pring(n)} \models \varphi$.

\begin{theorem}[\cite{Emerso03}, Theorem 3] \label{thm:EN}

  For interleaving parameterized token ring systems with no global inputs and parameterized specifications $\forall{i}.\pforall_{\forall{i}. ass(i)} \varphi(i)$, where $\varphi(i)$ and $ass(i)$ are \LTL\ formulas over process $i$, the cutoff is 2.

\end{theorem}

\begin{corollary}[\cite{Jacobs14}] \label{cor:en96forSynt}

  The parameterized synthesis problem of interleaving token rings with no global inputs for parameterized specifications $\forall{i}.\pforall_{\forall{i}. ass(i)} \varphi(i)$, where $\varphi(i)$ and $ass(i)$ are $\LTL$ formulas over process $i$, can be reduced to the synthesis problem of the token ring of size 2.

\end{corollary}
%
%
%
%

\subsection{Bounded Synthesis Method}\label{sec:bounded-synthesis}
By bounding the desired size of implementations, \emph{bounded synthesis}~\cite{Finkbeiner13} reduces the synthesis problem to a sequence of SMT problems. 
Uninterpreted functions are used to describe the transition relation, output functions, and auxiliary `ranking' functions, and the SMT solver tries to find valuations of these functions such that the specification is satisfied.
The flow is the following:
\begin{enumerate}
  \item \textbf{Automata translation}: 
  The negation of a given specification $\varphi$ is translated into a non-deterministic \Buchi\ automaton $A_{\neg\varphi}$.

  \item \textbf{SMT encoding}:  
  We encode a ranking function $\rank$ on states of the product of the specification automaton $A_{\neg\varphi}$ and the uninterpreted system. Consider a transition from composed state $(q,s)$ to $(q',s')$, where $q,q'$ are states of $A_{\neg\varphi}$ and $s,s'$ are states of the system. 
  Then we require $\rank(q',s')>\rank(q,s)$ if $q'$ is an accepting state of $A_{\neg\varphi}$, and otherwise $\rank(q',s')\geq\rank(q,s)$.
  The rule ensures that the SMT constraints are satisfiable if and only if the product does not have loops with an accepting state of the automaton, and a solution represents a correct implementation of the system.
  
  \item \textbf{SMT solving, iteration for increasing bounds}: If the SMT constraints in step $2$ are satisfiable, then return the implementation. Otherwise, there exists no implementation of the given size bound; we increase the bound and repeat step $2$.

\end{enumerate}
For the details of the SMT encoding that we use for synthesis of token rings see Khalimov et al.~\cite{TEPS}.

\section{Challenges for the Existing Approach}\label{sec:challenges}
Based on the existing approach for parameterized synthesis, we want to synthesize a system that satisfies 
$$ \pforall \left( (A1 \land \ldots \land A4 \land \FairSched) \rightarrow (G1 \land \ldots \land G11 \land TR) \right),$$
where
\begin{itemize}
\item 
$\FairSched$ is the form $\forall{i}.\always\eventually sch_i$, specifying that every process is scheduled infinitely often.
  \item 
 $TR$ are guarantees ensuring that the process template satisfies the requirements of the token ring process template defined in Sect.~\ref{def:process_template}: 
\[\begin{array}{lll}
\forall{i}. & \always (\sendi \impl \toki) \\
\forall{i}. & \always (\toki \land \neg \sendi \impl \nextt \toki) \\
\forall{i}. & \always (\neg\toki \land \neg \send[i-1] \impl \nextt \neg \toki) \\
\forall{i}. & \always (\toki \impl \eventually \sendi) 
\end{array}\]
\end{itemize}

As the existing cutoff results of Emerson and Namjoshi~\cite{Emerso03} (or their extensions by Aminof et al.~\cite{AminofJKR14}) do not support all features of the AMBA specification, we need to address the following challenges:

\begin{enumerate}
\item {\bf Synchronous AMBA and global inputs}:
        The AMBA protocol uses synchronous timing and has several global inputs (that are shared between all processes), while the  
cutoff results in \cite{Emerso03,AminofJKR14} are for interleaving systems with local action labels instead of inputs. We have discussed in Sect.~\ref{sec:semantics} how actions simulate local inputs, but global inputs are not supported in the existing cutoff theorems.
 
        In Sect.~\ref{sec:sync-case} we extend the cutoff results to fully asynchronous token rings with global inputs. For our synthesis approach, we will use the fact that correctness of an implementation in the fully asynchronous case implies correctness in the synchronous case.
		
        \item {\bf Global outputs}: The AMBA specification assumes that there 
are global outputs, i.e., those that depend on the global state of the system, such as \hmastlock. 
This is not handled by \cite{Emerso03,AminofJKR14}\ak{CANNOT in principle, or just not, or we could not do this?}, and in Sect.~\ref{sec:global-outputs} we address this by synthesizing local outputs that can be manually converted to suitable global outputs with simple logical operations.

        
                


        \item {\bf Special 0-process, immediate reaction, global information}: 
The AMBA specification distinguishes between master number 0 and all other masters.
We support this by synthesizing two different process implementations, one that serves master 0, and one for all other processes. Furthermore, process 0 is supposed to immediately grant master 0 when no process receives a \hbusreqi\ signal - this is a problem since only processes that have the token should give a grant, and information about requests of other processes is not available to process 0. We show how to handle this by weakening the specification and introducing an auxiliary global input in Sect.~\ref{sec:zero-process}.

\end{enumerate}


\section{Obtaining and Handling a Parameterized AMBA Specification}
\label{sec:extensions}

In the following, we will show how we obtained a parameterized AMBA specification suitable to our parameterized synthesis approach, and how we extended the approach to handle this specification.

\subsection{Addressing Challenges `Synchronous AMBA' and `Global Inputs'}
\label{sec:global-inputs}\label{sec:sync-case}

We will first extend the cutoff results for token rings to fully asynchronous systems with global inputs, for restricted classes of process templates and assumptions. Since these classes are not sufficient to model AMBA, we will afterwards introduce a method to localize assumptions in a sound but incomplete way.

\subsubsection{Complete Approach: New Cutoff Results} 

We consider systems and specifications that satisfy the following assumptions:
\begin{itemize}
  \item [a)]
$P= (\Locals, \LocalsI, \ActionsProc, \Trans, \LabelFun)$ is such that: $\forall{q \in \Locals}$ with $\tsnd \in \LabelFun(q)$ there exists unique ${q' \in \Locals}$ such that $\trans{q}{q'}{in}$ for any input ${\sf in} \in \ActionsProc$. I.e., in all sending states the process ignores inputs.\sj{here it should be enough to require this for global inputs}
  
  \item [b)] The assumptions $\forall{i}.ass(i)$ are of the form $\forall{i}.\always\alpha(i)$ or of the form $\forall{i}.\alpha(i)$, where $\alpha(i)$ is a Boolean formula over inputs (including global inputs) of process $i$.
\end{itemize}
Then (proofs can be found in the appendix):
\begin{theorem}\label{thm:main-theorem}
  Assume conditions (a) and (b). Then, for parameterized fully asynchronous token ring systems and parameterized specifications as stated below, the cutoffs are: 
  \begin{itemize}
    \item for $\forall{i}.\pforall_{\forall{i} \always\alpha(i)} \varphi(i)$ the cutoff is 2,
    \item for $\forall{i,j}.\pforall_{\forall{i} \always\alpha(i)} \psi(i,j)$ the cutoff is 4,
  \end{itemize}
  where 
  $\alpha(i)$ is a Boolean formula over inputs of process $i$, 
  $\varphi(i)$ is an \LTL\ formula over inputs and outputs of process $i$, 
  and $\psi(i,j)$ is an \LTLmX\ formula over outputs of processes $i,j$.

\end{theorem}

Note that the problem becomes undecidable if we do not restrict fair path properties, i.e., if we remove (b) but still assume (a): 
\begin{observation} \label{obs:undec}
 
  The parameterized model checking problem for fully asynchronous token rings with global inputs and properties of the form 
  $\forall{i}.\pforall_{\forall{i}.ass(i)} \varphi(i)$ 
  is undecidable, 
  where $ass(i)$ and $\varphi(i)$ are \LTL\ formulas over inputs and outputs of process $i$ (including global inputs).

\end{observation}

Note that Theorem~\ref{thm:main-theorem} does not support all assumptions in the AMBA specification (Fig.~\ref{fig:AMBASpec}): A3 and A4 are supported by the theorem, but A1 and A2 are not.

\subsubsection{Incomplete Approach: Localization of Assumptions}
Since Theorem~\ref{thm:main-theorem} does not support assumptions A1 and A2,
  we introduce an approach that \emph{localizes} the assumptions, essentially rewriting the specification $\forall j.\ \pforall_{\forall i.\ ass(i)} \varphi(j)$ into a form $\forall j. \pforall \left( ass(j) \rightarrow \varphi(j) \right)$.\footnote{Note that in some cases, the localized version is equivalent to the original one, e.g. for A2, since $\hready$ is a global input.} 

However, this naive form of localization strengthens the AMBA specification too much, making it unrealizable. Instead, we use a specialized way for localizing assumptions in token rings.

\paragraph{Localization of assumptions in token rings.} As suggested in \cite[Sect.6]{TEPS},
	$$\pforall_{\forall{i}.\ ass(i)} \forall{j}.\ (gua(j) \land TR(j))$$ 
  is localized into 
  $$\forall{i}.\ \pforall\ (ass(i) \impl TR(i)) \land (ass(i) \land \GF\toki \impl gua(i)),$$ where $ass(i)$ includes $\FairSched$, and $TR$ are the token ring properties as defined in Sect.~\ref{sec:challenges}.
	
  This restores realizability in our case. Intuitively, this specification guarantees that $TR$ will be satisfied under the given local assumptions, and for the rest of the guarantees we can then assume that all other processes will eventually send the token, thus satisfying the additional assumption $\GF\toki$.
	
\paragraph{Linking token possession to mutual exclusion.}
  In addition to global assumptions, the original specification contains an implicit mutual exclusion property: G4 defines how \hmaster\ is updated by the \hgranti\ signals. Note that G4 can only be satisfied if the \hgranti\ are mutually exclusive. 
	Since we know that the token can (and must) be used to ensure mutual exclusion, we explicitly specify this by adding G12: $\forall i.\,\hgranti \rightarrow \toki$. Together with localization of assumptions, this ensures that the parameterized specification will be $1$-indexed.
		
\paragraph{Resulting specification.}
The resulting specification is of the form 

\[ \forall i.\ \pforall \left( \left( ass(i) \impl TR(i) \right) \land \left( (ass(i) \land \GF\toki) \impl gua(i) \land G12 \right) \right), \]

i.e., a $1$-indexed LTL property in prenex-indexed form.
	  
	While Theorem~\ref{thm:main-theorem} supports some formulas of the type $\pforall_{\forall i.\ ass(i)} \forall j.\ gua(j)$, solving the synthesis problem for formulas with assumptions in this form is costly. In particular, every liveness assumption introduces a loop (with length equal to the size of the ring under consideration) for every liveness guarantee in the specification. This severely blows up the size of the specification automaton. Thus, even for the liveness assumptions A3 and A4 that are supported by the theorem, we use the \emph{localization} approach.

\subsection{Addressing Challenge `Global Outputs'}
\label{sec:global-outputs}

To address this challenge we define what is a localizable global output, introduce a special version of localizable global outputs we use for AMBA, and modify the specification to handle these global outputs. 

\paragraph{Linking global to local outputs.}
For a given parameterized system, a \emph{localizable global output} is a global output that can be expressed as a propositional formula over terms of the form $\forall{i}. \alpha(i)$ and $\exists{i}. \alpha(i)$, where $\alpha(i)$ is a propositional formula over outputs of process $i$.

\paragraph{Fixed solution for AMBA.}
  The AMBA specification in Fig.~\ref{fig:AMBASpec} has global outputs 
\hmastlock, \hstart, \hdecide, and \hmaster.
  We restrict synthesis to search for a solution with a fixed localizable implementation of global outputs, namely: For each global output signal $g$ we introduce a local output signal $g_i$, and define 
	\begin{itemize}
	\item $\hmaster := i$ whenever $\hmasteri$ is high\ak{doesn't fit the def}, and 
	\item $g:= \exists{i}.\ \toki \land g_i$ for all other global outputs $g$.
	\end{itemize}
  %
  
\paragraph{Modification of the parameterized specification.}
  According to the two previous steps, we should replace all global outputs in the specification with their specialized localizable definitions in terms of the new local outputs.
  For example, $\hstart$ should be replaced by $\exists{i}.\toki \land \hstarti$. 
  However, in token ring systems the only communication between processes is token passing, and hence the value of $\exists{i}.\toki \land \hstarti$ is not known to a process, except when it has the token (and thus defines that value).
%
  Thus, we replace each global output with its local version, e.g., $\hstart$ is replaced by $\hstarti$.
  
  Note that the limited communication interface (via token passing) does not make AMBA unrealizable, even though processes cannot access the value of global outputs when they do not possess. Intuitively, this is because the token is the shared resource that guarantees mutual exclusion of grants, and therefore the values of these global signals should always be controlled by the process that has the token. In particular, outputs \hdecide\ and \hstart\ are signals that are used to decide when to raise a grant and when to start and end a bus access\footnote{The original AMBA specification~\cite{AMBASpec} does not have these signals -- they were introduced to simplify the formalization of the specification~\cite{BarbaraThesis}.}, which should only be done when the token is present.
  Similarly, signals \hmastlock\ and \hmaster\ should be controlled by the process that currently controls the bus (and hence has the token). 
  By using only the local version of these signals in the specification, we force the implementation to never raise them unless the process has the token.

\subsection{Addressing Challenges `Special 0-process' and `Global Information'}\label{sec:zero-process}
The AMBA specification is of the form 
$\pforall_{\forall{i}.ass(i)}(\forall{i \neq 0}.\varphi(i) \land \psi(0))$, i.e., it distinguishes the behavior of process $0$. 
Recall the AMBA guarantees G10 from Fig.~\ref{fig:AMBASpec} (after localization steps of the previous sections):
\[\begin{array}{rrlr}
\forall \mathrm i\neq 0: & \always & \neg \hgranti \impl (\neg \hgranti \weakuntil \hbusreqi) & (G10.1)\\[3pt]
& \always & (\hdecide[0] \land (\forall \mathrm i: \neg \hbusreqi)) \impl \nextt \hgrant[0] & (G10.2) \\[3pt]
\end{array}\]
The distinction between $0$- and non-$0$-processes, as well as the required properties, present several additional challenges to the parameterized synthesis approach.

\paragraph{Distinguished $0$-process.}
The process templates for 0- and non-0-processes for specifications of the form 
$\pforall_{\forall{i}.ass(i)}(\forall{i \neq 0}.\varphi(i) \land \psi(0))$
can be synthesized separately, 
i.e., first, synthesize a process template $P_\varphi$ for 
$\pforall_{\forall{i}.ass(i)}\forall{i}.\varphi(i)$ 
and a process template $P_\psi$ for $\pforall_{\forall{i}.ass(i)}\forall{i}.\psi(i)$.
Then a combined token ring consisting of any number of copies of $P_\varphi$ and of one copy of $P_\psi$ at 0 vertex will satisfy $\pforall_{\forall{i}.ass(i)}(\forall{i \neq 0}.\varphi(i) \land \psi(0))$\ak{why?}. 
Hence we introduce a separate specification for the 0-process and synthesize it separately. 

To this end, we also separate G11 into two parts, G11.1: $\neg \hgranti \land \neg \hmastlocki$ (for non-$0$-processes) and G11.2: $\tok[0] \impl \hgrant[0] \land \hmaster[0] \land \neg \hmastlock[0]$ (for $0$-process).

%

\paragraph{Immediate reaction.}
Guarantee G10.2 requires an immediate reaction to a state where no process receives a bus request. 
This is unrealizable for AMBA in token rings because mutual exclusion of the grants requires possession of the token and implies $\always(\hgranti \impl \toki)$. 
To allow the process to wait for the token and then immediately react, we modify G10.2 to 
$\always \ (\neg \tok[0] \land \nextt \tok[0] \land \forall \mathrm i: \neg \hbusreqi)
\impl \nextt \hgrant[0])$
\footnote{You probably expect to see 
$\always \ (\hdecide[0] \land (\forall \mathrm i: \neg \hbusreqi) \land \nextt \tok[0])
\impl \nextt \hgrant[0])$, 
but this makes the specification unrealizable due to the token ring requirement 
$\always (\tok \impl \eventually \send)$: 
the environment falsifies it by making $\forall \mathrm i: \neg \hbusreqi$ true whenever the process raises $\hdecide$ (and hence the process should continue granting and cannot release the token).
To regain realizability one could add additional assumptions (something like $\always\eventually (\hdecide\land\neg\norequests)$). 
Instead, we decided to change slightly the semantics of the specification.
}.

\paragraph{Global information.}
  G10.2 contains an index quantifier $\forall{i}$ inside the temporal operator $\always$, which is not supported by Thm.~\ref{thm:main-theorem}.
%
Intuitively, G10.2 requires 0-process to have \emph{global information} about inputs of all processes, as it needs to react to a situation where \hbusreqi\ is low for all i. This is not possible when only \hbusreq[0] is available as an input, so we introduce an auxiliary (global) input $\norequests$, and add the assumption $\forall{i}.\always (\hbusreqi \impl \neg \norequests)$. 
Then G10.2 becomes:
$\always \ 
(\neg \tok[0] \land \nextt \tok[0] \land \norequests)
\impl \nextt \hgrant[0])$.
%
%
Such guarantees and assumptions are allowed by Thm.~\ref{thm:main-theorem}.


\subsection{Resulting Parameterized AMBA Specification} \label{sec:spec-translation}
\begin{figure}
  \fbox{%
  \begin{minipage}{\textwidth}
  \input{amba-spec-new}
  \end{minipage}
  }
  \caption{Parameterized AMBA specification for non-$0$-processes. G10.2 is only needed for 0-process.}
  \label{fig:AMBASpecNewI}
\end{figure}
%
\begin{figure}
  \fbox{%
  \begin{minipage}{\textwidth}
  \vspace{-10pt}
  \input{amba-spec-new0}
  \end{minipage}}
  \caption{Parameterized AMBA specification for $0$-process: modifications wrt. non-$0$-processes.}
  \label{fig:AMBASpecNew0}
\end{figure}
We obtained the new specification from the one in Fig.~\ref{fig:AMBASpec} by localization of global assumptions (Sect.~\ref{sec:global-inputs}), localization of global output signals \hmaster, \hmastlock, \hdecide, and \hstart\ (Sect.~\ref{sec:global-outputs}), and separation of specifications for $0$- and non-$0$-processes (Sect.~\ref{sec:zero-process}).
The resulting assumptions and guarantees for non-$0$-processes are given in Fig.~\ref{fig:AMBASpecNewI}, the modifications for the $0$-process in Fig.~\ref{fig:AMBASpecNew0}. The specifications to be synthesized are 
\[ \forall i.\ \pforall \left( \left( A1 \land \ldots \land A4 \impl TR(i) \right) \land \left( A1 \land \ldots \land A5 \impl G1 \land \ldots \land G10.1 \land G11.1 \land G12 \right) \right), \mathrm{~and}\]
\[ \forall i.\ \pforall \left( \left( A1 \land \ldots \land A4 \land A6 \impl TR(i) \right) \land \left( A1 \land \ldots \land A6 \impl G1 \land \ldots \land G10.2 \land G11.2 \land G12 \right) \right). \]
%
%
%
%


\section{Optimizations and Experiments}
\label{sec:optimizations}

In this section, we describe optimizations that proved to be crucial for the synthesis of the parameterized AMBA AHB, and present the results of parameterized synthesis in form of runtimes and resulting component implementations.

\paragraph{Prototype.}
The basis of our experiments is \textsc{Party}, a tool for parameterized synthesis of token rings~\cite{Party}. 
\textsc{Party} is written in Python, uses LTL3BA~\cite{ltl3ba} for automata translation and Z3~\cite{Moura08} for SMT solving.
All experiments were done on a x86\_64 machine with $2.60$GHz, $12$GB RAM.
Prototype implementation and specification files can be found at \url{https://github.com/5nizza/Party/} (branch `amba-gr1').

\paragraph{Synchronous Hub Abstraction \cite[Sect.6]{TEPS}.}
Synchronous hub abstraction can be applied to 1-indexed specifications. 
It lets the environment simulate all but one process, and always schedules this process.
Thus, the synthesizer searches for a process template in synchronous setting with additional assumptions on the environment, namely: i) the environment sends the token to the process infinitely often, and ii) the environment never sends the token to the process if it already has it.
Note that synchronous hub abstraction is sound and complete for the semantics of 1-indexed properties introduced in Sect.~\ref{sec:semantics}.
Also note that after applying this optimization any monolithic synthesis method can be applied to the resulting specification in Sect.~\ref{sec:spec-translation}.

\paragraph{Hardcoding States With and Without the Token~\cite[Sect.4]{TEPS}.}
The number of states with and without the token in a process template defines the degree of the  parallelism in a token ring. 
Parallelism increases with the number of states that do not have the token.
In the AMBA case study, any action with grant depends on having the token.
Thus we divide states into one that does not have the token, and all others that have the token, by hardcoding the $\toki$ output function.

\begin{figure}[t]\centering{
\includegraphics[width=\textwidth]{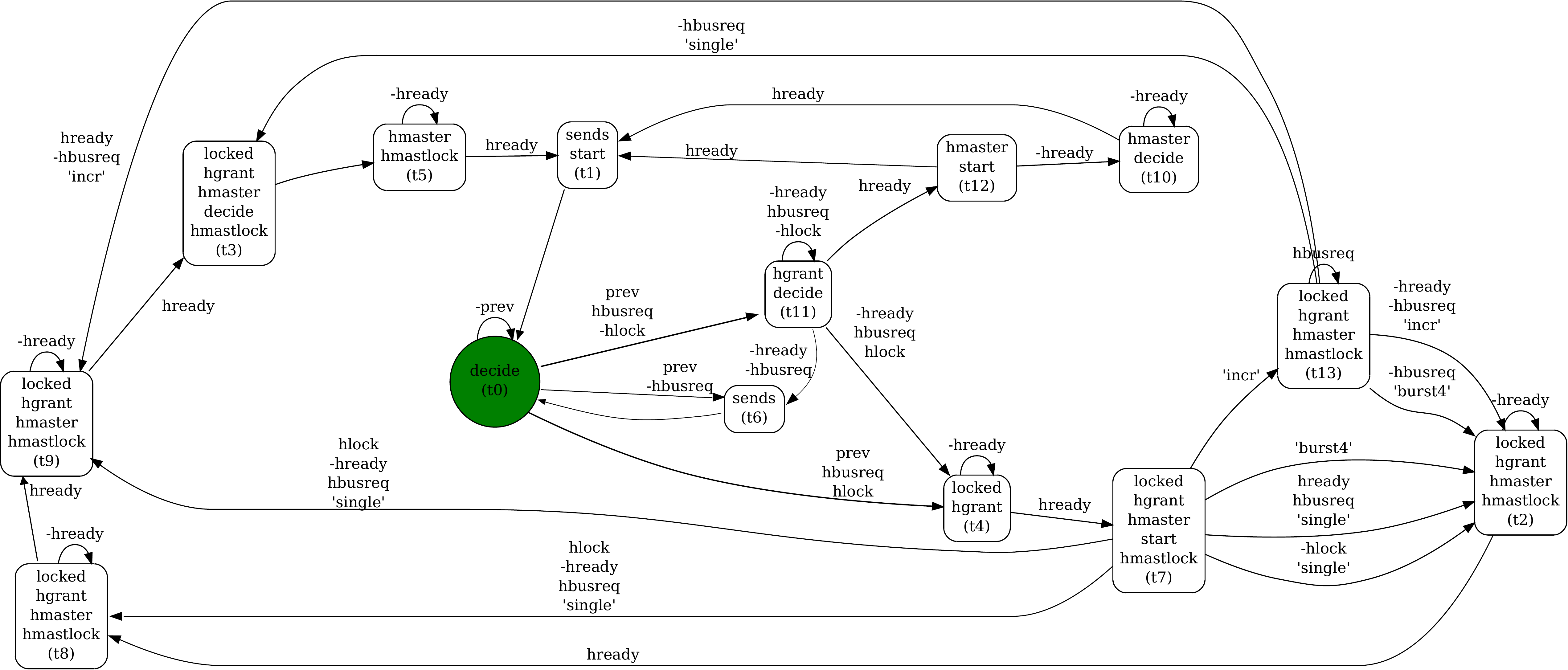}}
\caption{Synthesized model of non-$0$-processes (after manual simplification). 
Circle green state ($t0$) is without the token, other states are with the token. 
Initial states are $t0,t1$.
States are labeled with their active outputs. 
Edges are labeled with inputs, a missing input variable means ``don't care''.
`Burst4' means $\hburst=\hburstfour$, `incr' means $\hburst=\hincr$, `single' means neither of them.
In the first step of decompositional synthesis states $t0,..,t9$ were synthesized, in the second $t10,..,t12$ were added, in the final step state $t13$ was added.}
\label{fig:ith-model}
\end{figure}

\paragraph{Decompositional Synthesis of Different Grant Schemes.}
The idea of the decompositional synthesis is: synthesize a subset of the properties, then synthesize a larger subset using the model from the previous step as basis. 
Consider an example of the synthesis of the non-0-process of AMBA. 
The flow is: 
\begin{enumerate}
  \item   
  Assume that every request is locked with \hburstfour, i.e., 
  add the assumption $\always(\hlocki\land\hburst=\hburstfour)$ to the specification.
  This implicitly removes guarantee G2 and assumption A1 from the specification.
  Synthesize the model. The resulting model has $10$ states (states $t0,..,t9$ and transitions between them in Fig.~\ref{fig:ith-model}).

  \item
  Use the model found in the previous step as a basis: assert the number of states, values of output functions in these states, transitions for inputs that satisfy the previous assumption. 
  Transitions for inputs that violate the assumption from step 1 are not asserted, and thus left to be synthesized.
  
  Now relax the assumptions: allow locked and non-locked \hburstfour\ requests, i.e., replace the previous assumption with $\always(\hburst=\hburstfour)$. 
  Again, this implicitly removes G2 and A1.
  In contrast to the last step, now guarantee G3 is not necessarily `activated' if there is a request. 

  Synthesize the model. 
  This may require increasing the number of states (and it does in the case of non-0 process) -- add new states and keep assertions on all the previous states.

  \item
  Assert the transitions of the model found, like in the previous step.
  
  Remove all added assumptions and consider the original specification.
  Synthesize the final model.

\end{enumerate}
Although for AMBA this approach was successful, it is not clear how general it is.
For example, it does not work if we start with locked \hburstfour\ and \hready\ always high, and then try to relax it.
Also, the separation into sets of properties to be synthesized was done manually.

\begin{figure}[t]\centering{
\includegraphics[width=\textwidth]{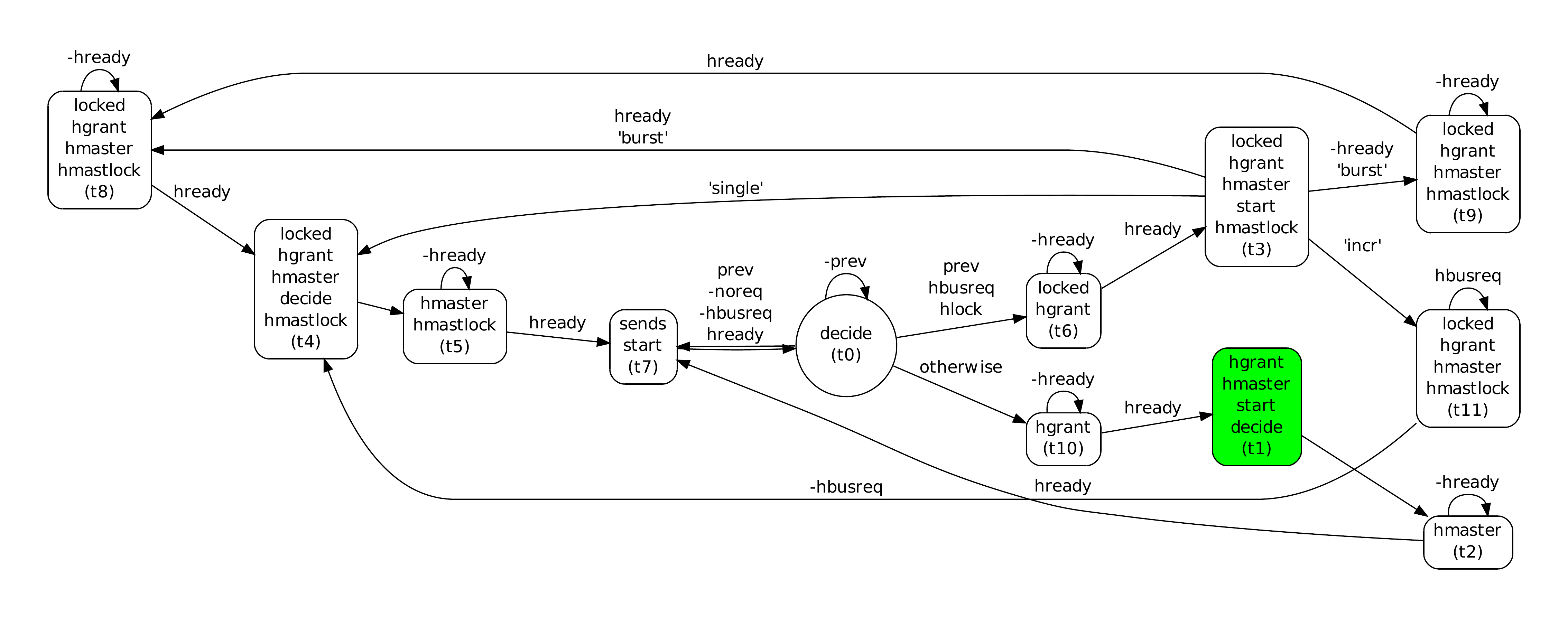}}
\caption{Synthesized model of $0$-processes (after manual simplification). 
Circle state ($t0$) is without the token, other states are with the token. 
Initial states are $t0,t1$.
States are labeled with their active outputs. 
Edges are labeled with inputs, a missing input variable means ``don't care''.
`Burst' means $\hburst=\hburstthree$ (recall we decreased the length of bursts for 0 process), `incr' means $\hburst=\hincr$, `single' means neither of them.
In the first step of decompositional synthesis states $t0,..,t10$ were synthesized, in the second only transitions were synthesized, but no new states added, in the final step $t11$ was added.}
\label{fig:zero-model}
\end{figure}
\todo{can we reuse ith models for the zero-process?}

\paragraph{Optimization of SMT Encoding.}\label{sec:simpleGR1}
\newcommand{\Atm}{\ensuremath{{A_{\neg\varphi}}}}
Recall from Sect.~\ref{sec:approach} that SMT based bounded synthesis, given an automaton \Atm\ of the negation of specification $\varphi$ and an unknown process template $P=(\Locals_P, \LocalsI, \ActionsProc, \Trans_P, out)$ with a fixed number of states, encodes the product automaton $\Atm \times P$ into SMT constraints such that $\Atm \times P$ contains no reachable loops with an accepting state of the \Atm\ iff SMT constraints are satisfiable.
Below is a general assertion from which the SMT query is composed: 
$$
\bigwedge{q\in Q_P}
\bigwedge{\trans{a}{b}{i,o}\in \delta_\Atm}: \ \ \ 
\rho(a,q) \ge 0 \land o=out(q) 
\impl
\rho(b,\delta_P(q,i)) 
%
\vartriangleright
\rho(a,q)
$$
where $\vartriangleright$ is `$>$' if $b$ is an accepting state of \Atm, else `$\ge$'.
In words: for any state of the process template, and any transition of the automaton, if the current state of the product automaton is reachable, then the next state should also be reachable and the ranking function should be as stated.

The specification of AMBA we synthesize is derived from GR(1) specification.
As a consequence it contain assumptions (A3, A6) of the form
$\always\alpha(i)$ where $\alpha(i)$ is a Boolean formula over current inputs, and many guarantees (G1, G4, G5, G6, G7, G8, G12, G10.2) of the form $\always\beta(i,o,o')$ where $\beta(i,o,o')$ is a Boolean formula over current inputs and outputs and next outputs.
Instead of using the standard approach via automaton translation described above, we:
\begin{enumerate}
  \item encode assertions of the form $\always\alpha(i)$ directly into SMT constraints, namely add $\alpha(i)$ to the the premise of the SMT rule. 
  Thus, the premise becomes  
  `$\rho(a,q)\ge 0 \land o=out(q) \mathbf{\land \alpha(i)} \impl ...$'

  \item for all guarantees of the form $\always\beta(i,o,o')$ add SMT constraints of the form: 
  $$
  \bigwedge{q \in Q_P} \bigwedge{i \in \ActionsProc}: \ \ 
  \alpha(i) \impl \beta(i,out(q),out(\delta_P(q,i)))
  $$
\end{enumerate}
The first optimization is sound and complete, the second one introduces incompleteness.

For AMBA specification in Fig.~\ref{fig:AMBASpecNewI} and \ref{fig:AMBASpecNew0} this optimization means that only guarantees G2, G3, G9, G10.1, G11 require the standard flow via automata translation.

Does this optimization help in the synthesis? 
Preliminary experiments (considering the first step of the decompositional synthesis of non-0 process) show: 
\begin{itemize}
\item 
With the optimization the automaton for the negated specification has 24 states, without -- 42 states.

\item
The synthesis time with optimization is 16 minutes, without -- 57 minutes.
Interesting to note that the optimized and non-optimized versions spent the same time (2 minutes) checking satisfiability of the last query (with the model size of 10), so the main difference is in checking unsatisfiable queries -- Z3 identifies unsatisfiability of optimized queries faster (14 vs. 53 minutes).
A similar behavior happens for a version of the same specification with reduced lengths of bursts ($3/4\impl 2/3$): total times are 3/6 minutes, but the last query took 1m40s/30s for optimized/non-optimized version. 

\end{itemize}

\paragraph{Results.}
Synthesis times are in Tables \ref{tab:non-zero-process} and \ref{tab:zero-process}, 
the model synthesized for non-0-process is in Fig.~\ref{fig:ith-model}.
The table has timings for the case when all optimizations described in this section are enabled --- it was not our goal to evaluate the optimizations separately, but to find a combination that works for the AMBA case study.

For the $0$-process we considered a simpler version with burst lengths reduced to 2/3 instead of the original 3/4 ticks. 
With the original length the synthesizer could not find a model within 2 hours (it hanged checking 11 states models while the model has at least 12 states).

Without the decompositional approach, the synthesizer could not find a model for for non-0 process of the AMBA specification within (at least) 5 hours.
\begin{table}
\centering
\begin{minipage}[b]{0.45\textwidth}
\caption{Results for non-$0$-process.}
\label{tab:non-zero-process}
\centering
\begin{tabular}{ l|cc }
Additional assumptions                          & time & \#states \\
\hline
\rule{0pt}{3ex} \specialcell{$\always \hlock$ \\ $\always \hburst=\hburstfour$} 
                       & 16min.  & 10  \\
\hline
\rule{0pt}{2ex} \specialcell{$\always \hburst=\hburstfour$} & 13sec.  & 13  \\
\hline
\rule{0pt}{2ex} 
-- (Full Specification) & 1min.  & 14 \\
\end{tabular}
\end{minipage}
\hspace{0.5cm}
\begin{minipage}[b]{0.45\textwidth}
\caption{Results for $0$-process \\ (bursts reduced: $3/4 \rightarrow 2/3$).}
\label{tab:zero-process}
\centering
\begin{tabular}{ l|cc }
Additional assumptions & time & \#states \\
\hline
\rule{0pt}{3ex} \specialcell{$\always \hlock$ \\ $\always \hburst=\hburstfour$} 
                       & 3h.  & 11  \\
\hline
\rule{0pt}{2ex} \specialcell{$\always \hburst=\hburstfour$} & 1min.  & 11  \\
\hline
\rule{0pt}{2ex} 
-- (Full Specification) & 1m30s.  & 12 \\
\end{tabular}
\end{minipage}
\end{table}
\ak{`decide' is not faithfully simulated in token passing simulations?}

\section{Conclusions}
\label{sec:conclusion}
\ak{mention VMCAI paper in the context: `one can see the token as a resource which gets granted, the token ring is a particular strategy for scheduling the access to the resource. But the models we synthesized work for any fair resource strategy not just token rings. This is true because of we have one-indexed properties and the abstraction completely destroys the topology information -- informal but smth like this. '}
We have shown that parameterized synthesis in token rings can be used to 
solve benchmark problems of significant size, in particular the well-known 
AMBA AHB specification that has been used as a synthesis benchmark for a long 
time. To achieve this goal, we extended slightly the cutoff results that 
parameterized synthesis is based on, and used a number of optimizations in 
the translation of the specification and the synthesis procedure itself to 
make the process feasible.

This is the first time that the AMBA case study, or any other realistic case 
study of significant size, has been solved by an automatic synthesis 
procedure for the general, parametric case. However, some of the steps in the 
procedure are manual or use an ad-hoc solution for the specific problem at 
hand, like the limited extension of cutoff results for global inputs, the 
construction of suitable functions to convert local to global outputs, or the 
decompositional synthesis for different grant schemes. Generalizing and 
automating these approaches is left for future work.

Furthermore, our synthesized implementation is such 
that the size of the parallel composition grows only linearly with the number of 
components. Thus, for this case study our approach does not only solve the 
problem of increasing synthesis time for a growing number of components, but 
also the problem of implementations that need an exponential amount of memory 
in the number of components. We pay for this small amount of memory with a 
less-than-optimal reaction time, as processes have to wait for the token in 
order to grant a request. This restriction could be remedied by extending the 
parameterized synthesis approach to different system models, e.g., processes 
that coordinate by guarded transitions, or communicate via broadcast messages.

\paragraph{Acknowledgments.} We thank Sasha Rubin for insightful discussions on cutoff results in token rings.

\bibliographystyle{eptcs}
\bibliography{synthesis,others,crossref}

\section{Appendix}

\subsection{Proof of Theorem~\ref{thm:main-theorem}}
\begin{proof}[Proof idea.]\ak{sync with def of param specs}
Consider the second case, the first one is similar.
For an arbitrary process template $P$ that satisfies (a), we need to prove that:
$$\forall{n\ge c}. P^{\pring(n)} 
\models 
\forall{i,j}.\pforall_{\forall{i}.ass(i)} \psi(i,j)
~\iff~
P^{\pring(c)} \models 
\forall{i,j}.\pforall_{\forall{i}.ass(i)} \psi(i,j).$$
The $\implies$ direction is trivial since the left part includes $P^{R\pring(c)}$.
The $\impliedby$ direction can be rewritten as 
$$\exists{n \ge c}. P^{\pring(n)} \models 
\exists{i,j}.\pexists_{\forall{i}.ass(i)} \neg\psi(i,j) 
\implies 
P^{\pring(c)} \models 
\exists{i,j}.\pexists_{\forall{i}.ass(i)} \neg\psi(i,j).$$
The proof uses the notion of stuttering bisimulation~\cite{bisimulation} and consists of two ideas: symmetry argument and simulation construction.

\noindent\textbf{Symmetry argument}:
To prove the cutoff results for
$\exists{i,j}.\pexists_{\forall{i}.ass(i)} \neg\psi(i,j)$
it is enough to prove it for
$\exists{j}.\pexists_{\forall{i}.ass(i)} \neg\psi(0,j)$.

Suppose that $\exists{i,j}.\pexists_{\forall{i}.ass(i)} \neg\psi(i,j)$ is satisfied in a token ring of size $n$. 
Consider $i,j \in \{0,\ldots,n-1\}$ that satisfy $\pexists_{\forall{i}.ass(i)} \neg\psi(i,j)$, and assume for simplicity $i<j$.
Consider a rotated token ring: process $0$ in the rotated ring repeats the behavior of process $i$ in the original token ring, process $1$ repeats the behavior of process $i\oplus 1$, and so on.
Then, the rotated ring will satisfy the property $\pexists_{\forall{i}.ass(i)} \neg\psi(0,j-i)$.

\noindent\textbf{Simulation construction}: 
if there is a run of a large system that satisfies 
$\exists{j}.\pexists_{\forall{i}.\always\alpha(i)} \neg\psi(0,j)$
then there is a run of the cutoff system that satisfies it.

For any $k\in\{2,..,n-2\}$:
$$
\exists{j}.\pexists_{\forall{i}.\always\alpha(i)} \neg\psi(0,j) 
\iff 
\pexists_{\forall{i}.\always\alpha(i)} \neg\psi(0,1) 
\lor 
\pexists_{\forall{i}.\always\alpha(i)} \neg\psi(0,k) 
\lor
\pexists_{\forall{i}.\always\alpha(i)} \neg\psi(0,n-1).
$$
Intuitively, the statement above holds because in a fully asynchronous token 
ring, a 2-indexed property can only `identify' whether the processes are 
direct neighbors or not, but not the number of processes in between.
Thus, if $\pexists_{\forall{i}.\always\alpha(i)} \neg\psi(0,k)$ holds for 
some $k\in\{2,\ldots,n-2\}$, then it holds for any $k\in\{2,\ldots,n-2\}$.
Let $k=2$, and thus consider
$$
\exists{j}.\pexists_{\forall{i}.\always\alpha(i)} \neg\psi(0,j) 
\iff 
\pexists_{\forall{i}.\always\alpha(i)} \neg\psi(0,1) 
\lor 
\pexists_{\forall{i}.\always\alpha(i)} \neg\psi(0,2) 
\lor
\pexists_{\forall{i}.\always\alpha(i)} \neg\psi(0,n-1).
$$

We will analyze the case where $\pexists_{\forall{i}.\always\alpha(i)} \neg\psi(0,2)$ is satisfied in the large ring. Other cases are similar.

We show how processes $0$ and $2$ in the cutoff ring (of size $4$) simulate transitions of corresponding processes of the ring of size $5$, see Fig.~\ref{fig:simulation}.
We assume for simplicity that process $0$ starts with the token.
A similar construction works the in general case for a ring of size $n$.
The construction, given a run of the large ring that 
satisfies ${\forall{i}.\always\alpha(i)}$ and $\neg\psi(0,2)$, builds a run of the cutoff ring that satisfies them:
\begin{itemize}
\item Until `moment 1' the system run of the large ring is repeated exactly in the cutoff ring.
\item At `moment 1', process 3 of the large token ring receives the token -- in the cutoff ring there is no corresponding process.
Hence, we stutter process $2$ in the cutoff ring in the state with the token, while other processes move as they move in the large token ring from moment 1 to 2.
\item Between moments 1 and 2, in the large ring process 2 may move without the token.
To simulate these transitions, in the cutoff ring process 2 sends the token to process 3 and then repeats the transitions of process 2 of the large ring between moments 1 and 2, while all other processes stutter.
We need to repeat these transitions to ensure the states of all processes of the cutoff ring equal to states of the corresponding processes in the large ring (in the beginning of moment 1)\footnote{Strictly speaking, the exact same global state (if we ignore the abstracted process) is not required for proving the theorem.}.
Thus, the path between moments 1 and 2 may be longer in the cutoff ring.
\item The problematic step in the figure is at moment 2. 
In the large ring, process 2 sends the token to process 3 and reads a `red' global input, while in the cutoff ring the construction provides a (possibly) different `black' input to the other processes. 
This is where we use the additional restrictions (a) and (b) on process template and formulas for fair paths -- they allow replacement of `red' with `black' input, such that (a) the behavior of the process does not change, and (b) the property will still be satisfied.
\item Finally, from moment 2 until process $0$ receives the token, the cutoff ring repeats transitions of the large token ring exactly. Then, the new round starts and we repeat the construction.
\end{itemize}

\emph{Correctness}.
The construction ensures: 
i) local runs\footnote{Recall that local run of process $i$ considers only the elements of the system run in which process $i$ moves, see Sect.~\ref{sec:preliminaries}.} of all the processes of the cutoff ring satisfy assumptions $\forall{i}.\always\alpha(i)$ if they were satisfied in the large ring, 
ii) global run, projected on outputs of processes 0 and 2 in the cutoff ring, is equal up to stuttering to that of processes 0 and 2 in the large ring\footnote{The construction does not ensure that global runs projected on outputs \emph{and} inputs are the same if consider inputs in all moments of the system run, i.e., even when processes $i,j$ do not move.}.
These two items imply that the system run of the cutoff ring satisfies ${\forall{i}.ass(i)}$ and $\neg\psi(0,2)$ if the system run of the large ring satisfies ${\forall{i}.ass(i)}$ and $\neg\psi(0,2)$.
\begin{figure}[t]\centering{
\includegraphics[scale=1.3]{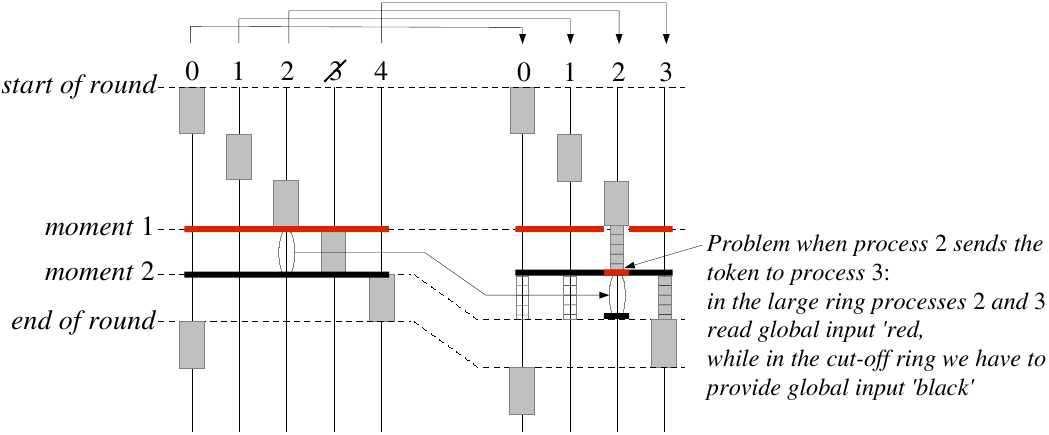}}
\caption{Simulating a system run of the large token ring of size 5 by a cutoff token ring of size 4. 
Behavior of processes 0,1,2,4 in the large ring is simulated by processes 0,1,2,3 respectively in the cutoff ring (behavior of process $3$ of the large token ring is abstracted away).
The run starts at the top, and evolves downwards.
Vertical solid lines denote process states (exact states are omitted).
Gray boxes denote states with the token, lined boxes mean stuttering (processes do not move).
Bold red and black horizontal lines denote different global inputs the processes received at this step.
}
\label{fig:simulation}
\end{figure}
\end{proof}

\subsection{Proof of Observation~\ref{obs:undec}}
\begin{proof}[Proof idea.]
In \cite[Section 6]{Emerso03} the authors prove that if the token is allowed to carry a single bit of information, then the parameterized model checking (even when a process template has no inputs) problem becomes undecidable. 
Global inputs and fair path assumptions can be used to simulate token values (which implies the undecidability of the parameterized model checking) in the following way: 
\begin{itemize}
  \item
  The process template has two states $snd_0,snd_1$ that send the token: 
  $\tsnd\in \LabelFun(snd_0) \land \tsnd\in \LabelFun(snd_1)$

  \item 
  $ass(i):=
  \always(snd_0 \impl glob_0) 
  \land 
  \always(snd_1 \impl glob_1) 
  \land 
  \always\neg(glob_0\land glob_1)
  $, 
  i.e., only paths on which the environment provides global input $glob_i$ when process sends the token from state $snd_i$ are considered.

\end{itemize}
This process template ensures: on paths that satisfy $\forall{i}.\always ass(i)$, if a process receives the token and reads global input $glob_i$, then the previous process sent the token from state $snd_i$.
This way a (value-less) token ring with global inputs and special assumptions on fair paths can model rings with the binary token.
\end{proof}

\end{document}